\documentclass[aps, prx, twocolumn, showpacs, floatfix, superscriptaddress, groupedaddress]{revtex4-1}
\usepackage[utf8]{inputenc}
\usepackage{amsthm}
\usepackage{amssymb}
\usepackage{amsmath}
\usepackage{gensymb}
\usepackage{graphicx}
\usepackage{url}
\usepackage{framed}
\usepackage{float}
\usepackage{wrapfig}
\usepackage{multirow} 
\usepackage{tikz}
\usepackage{fullpage}
\usepackage[toc,page]{appendix}
\usepackage{physics}
\usepackage{listings}
\usepackage{hyperref}
\hypersetup{
    colorlinks=true,
    linkcolor=blue,
    citecolor=blue,
    filecolor=magenta,      
    urlcolor=blue,
}

\usepackage{natbib}

\usetikzlibrary{arrows}
\usetikzlibrary{decorations.markings}

\usepackage{csquotes}
\MakeOuterQuote{"}

\begin{document}


\title{Revisiting Coulomb diamond signatures in quantum Hall interferometers}


\author{N. Moreau$^{1\dagger}$, S. Faniel$^{2\dagger}$, F. Martins$^{1\sharp}$, L. Desplanque$^{3}$, X. Wallart$^{3}$, S. Melinte$^{2}$, V. Bayot$^1$, \& B. Hackens$^{1*}$}

\affiliation{$^1$IMCN/NAPS, Universit\'e catholique de Louvain, B-1348 Louvain-la-Neuve, Belgium \\
$^2$ICTEAM, Universit\'e catholique de Louvain, B-1348 Louvain-la-Neuve, Belgium\\
$^3$University Lille, CNRS, Centrale Lille, JUNIA ISEN, University Polytechnique Hauts de France, UMR 8520-IEMN F-59000 Lille, France\\
$\dagger$ These authors contributed equally to this work\\
$^{*}$ corresponding author : benoit.hackens@uclouvain.be\\
$^{\sharp}$ Present address: Hitachi Cambridge Laboratory, J.J. Thomson Avenue, Cambridge CB3 0HE, United Kingdom}

%

\date{\today}

\begin{abstract}
Coulomb diamonds are the archetypal signatures of Coulomb blockade, a well-known charging effect mainly observed in nanometer-sized "electronic islands" tunnel-coupled with charge reservoirs. Here, we identify apparent Coulomb diamond features in the scanning gate spectroscopy of a quantum point contact carved out of a semiconductor heterostructure, in the quantum Hall regime. Varying the scanning gate parameters and the magnetic field, the diamonds are found to smoothly evolve to checkerboard patterns. To explain this surprising behavior, we put forward a model which relies on the presence of a nanometer-sized Fabry-Pérot quantum Hall interferometer at the center of the constriction with tunable tunneling paths coupling the central part of the interferometer to the quantum Hall channels running along the device edges. Both types of signatures, diamonds and checkerboards, and the observed transition, are reproduced by simply varying the interferometer size and the transmission probabilities at the tunneling paths. The new proposed interpretation of diamond phenomenology will likely lead to revisit previous data, and opens the way towards engineering more complex interferometric devices with nanoscale dimensions.
\end{abstract}

\maketitle

\section{Introduction}
When a two-dimensional electronic gas (2DEG) is placed in a large perpendicular magnetic field, charge carriers flow in one-dimensional quantum Hall edge channels (QHECs) formed as the Landau levels (LLs) cross the Fermi energy along device edges. If charge carrier phase coherence is preserved over sufficiently long distances, these edge channels can play the role of monochromatic wave beams to form quantum Hall interferometers (QHIs), the counterparts of optical interferometers in electronic systems. QHI device design relies on beam-splitters bringing the interfering QHECs in close proximity to modulate the reflection $r$ and the transmission $t=1-r$ of charge carriers in and out the interferometer \cite{Halperin2011}. In this framework, Fabry-Perot (FP) \cite{Wees1989} and Mach-Zehnder \cite{Ji2003} QHIs have been successfully implemented, displaying clear interference due to the Aharonov-Bohm (AB) effect. They offer a promising path towards quantum computing based on anyonic braiding \cite{Baeuerle2018,Nakamura2020,Bartolomei2020} as well as opportunities to test the groundwork of quantum physics, e.g. through tests of Bell inequalities violation in an Hanbury Brown and Twiss configuration \cite{Samuelsson2004}. Note that the latter objective can only be reached provided that QHIs have a small size compared with the phase coherence length \cite{Neder2007}.

Nevertheless, small QHIs are plagued by Coulomb charging effects, which mask AB interferences. Indeed, Coulomb blockade may also lead to oscillations of the QHI conductance when varying the magnetic flux \cite{Zhang2009,kou2012}. Inferring the precise origin of such oscillations, and hence the regime of the studied QHI, is achieved via spectroscopy measurements. They consist in measuring the conductance through the QHI while varying both the DC bias and the magnetic flux. If the resulting map draws a checkerboard pattern, it indicates that the QHI operates in the AB regime \cite{Neder2006,Roulleau2007,McClure2009,Martins2013,Nakamura2019} whereas a diamonds pattern is associated to the Coulomb dominated (CD) regime in the literature \cite{Kataoka1999,McClure2009,Martins2013a}. 

Such spectrosopies, combined with scanning gate microscopy (SGM), have revealed that a QHI can form spontaneously in the vicinity of a quantum point contact (QPC) where an antidot, associated to potential inhomogeneities, couples to the counterpropagating QHECs flowing along both sides of the constriction \cite{Martins2013,Martins2013a}. Intriguingly, these QHIs have been found to exhibit both the AB \cite{Martins2013} and the CD signatures \cite{Martins2013a}.

Here, we study a semiconductor-based 2DEG patterned in a QPC geometry in the quantum Hall (QH) regime. Thanks to SGM, we reveal the presence of a natural nanometer-sized QHI, associated with an impurity-induced quantum dot located in the vicinity of the constriction. Spectroscopy on this QHI exhibits a continuous evolution from checkerboard to diamond patterns when varying the magnetic flux. With the help of a basic FP model, we show that the transition between both types of signatures can be explained \textit{without} invoking Coulomb charging effects. Instead, it can be reproduced assuming a smooth change of the reflection and transmission amplitudes ($r$, $t$).

\section{Results}
\subsection{Experimental setup}
\begin{figure}[!ht]
\centering
\includegraphics[width=\linewidth]{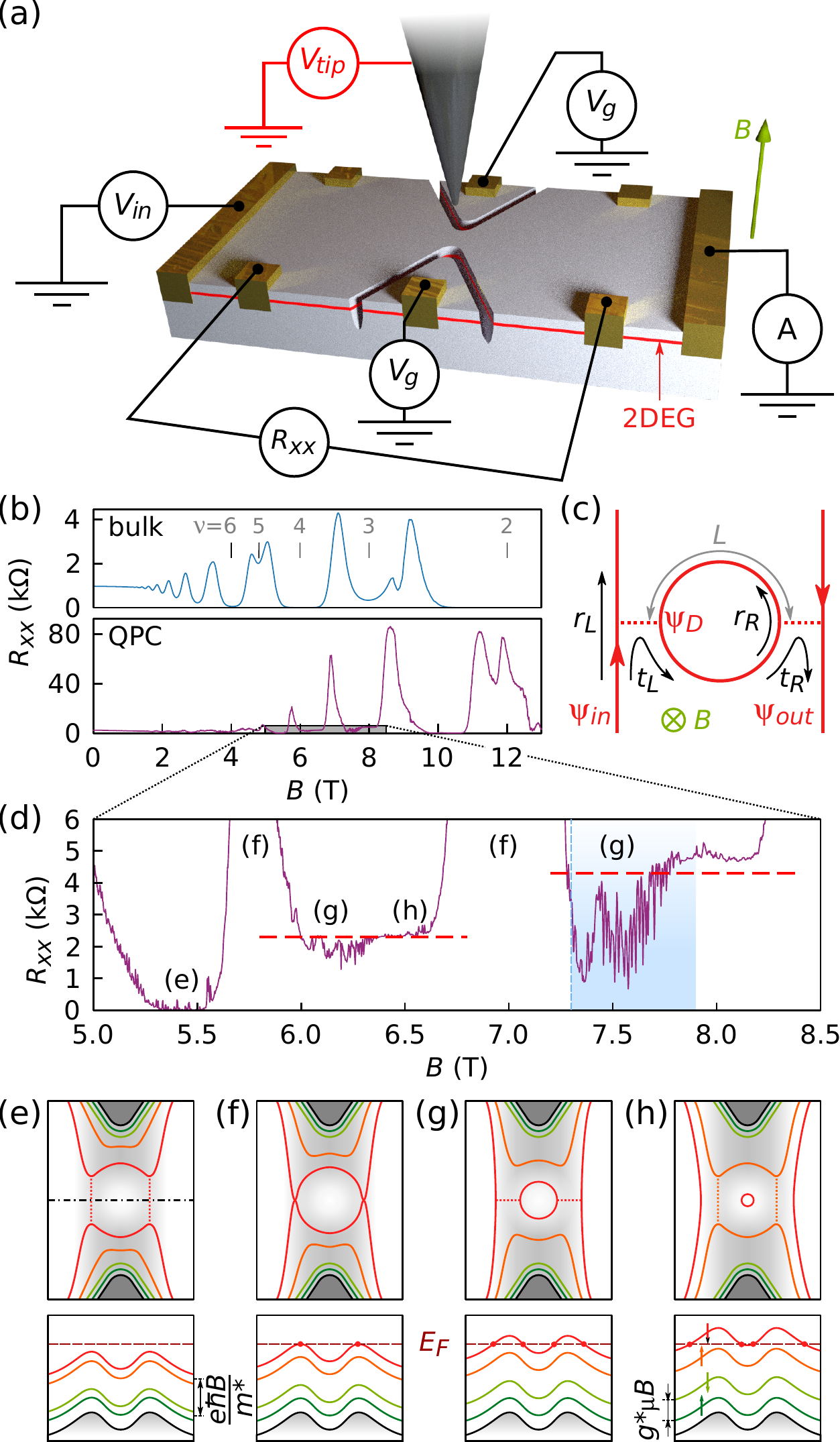}
\caption{(a) Artist view of the experimental setup with the buried 2DEG depicted in red. The biased tip locally changes the electron density when applying the voltage $V_{tip}$ and is moved at a distance $d_{tip} \sim 60$ nm above the 2DEG. A magnetic field $B$ is applied perpendicularly to the 2DEG plane. (b) $R_{xx}$ as a function of $B$ measured on a Hall bar patterned next to the QPC (top panel) and in the QPC (bottom panel) with $V_g = 5$ V. The vertical lines indicate integer filling factors. (c) Schematic of a QHI formed by a QHEC loop coupling two counterpropagating QHECs (the dotted lines correspond to tunneling paths). (d) Zoom of the gray shaded region in (b). $2.2 ~ {\rm k\Omega}$ and $4.3 ~ {\rm k\Omega}$ plateaus are indicated with red dashed lines. (e-h) Evolution of the QHECs spatial configuration with $B$ in the vicinity of the constriction (top panels) and the corresponding Landau level positions compared to the Fermi energy $E_F$. Higher potential (lower electron density) is depicted with darker gray shades.}
\label{figure1}
\end{figure}

The device is built from an $\rm In_{0.7}Ga_{0.3}As/In_{0.52}Al_{0.48}As$ quantum well grown by molecular beam epitaxy on an InP substrate (see \cite{Toussaint2018,Toussaint2018a}). A 15 nm-thick quantum well buried 25 nm below the surface hosts a 2DEG with a bare electronic density $N_S = 5.7 \cdot 10^{15} \ \rm m^{-2}$ and a mobility $\mu = 5.3 \cdot 10^4 \ \rm cm^2/Vs$. The 2DEG was electrically contacted by means of  Ge/Au ohmic contacts. The Hall bar and the QPC were patterned using electron beam lithography and wet etching. The QPC materializes thanks to two narrow trenches defining a $\sim 350 \ \rm nm$-wide constriction (see Fig. \ref{figure1}a). The 2DEG areas beyond the etched trenches of the QPC were used as side gates (with an applied bias $V_g$) in order to control the effective width of the constriction. 

Transport measurements were performed in a dilution refrigerator (base temperature $T \approx 100 \ \rm mK$) with a magnetic field $B$ applied perpendicular to the plane of the 2DEG. The sample's resistance $R_{xx}=dV/dI$ was measured using a standard ac lock-in techniques with a 2 nA excitation current, and frequencies between 8 Hz and 90 Hz. The dilution refrigerator is also equipped with an atomic force microscope whose metallic tip was used as a local gate. Using this microscope, we acquired SGM images of the device by recording its resistance while the tip biased at a voltage $V_{\rm tip}$ was scanning the sample's surface at a distance of $\sim 60$ nm. In addition, two types of spectroscopies were obtained. For this purpose, a DC electric current $I_{sd}$ was added to the lock-in signal, and the QPC's differential resistance $R_{xx}=\left. dV/dI \right|_{I_{sd}} $ was recorded as a function of $I_{sd}$ and $B$ or $I_{sd}$ and $V_{\rm tip}$, keeping the tip at a fixed position.

\subsection{Evolution of the magnetoresistance}\label{magneto}
Figure \ref{figure1}b compares the evolution of $R_{xx}$ as a function of $B$ with $V_g=5 ~ \rm V$ in two different regions of the sample : a macroscopic Hall bar (top panel) and the QPC region (bottom panel). Both curves show Shubnikov-de Haas oscillations and QH effect stemming from the magnetic depopulation of the LLs in the 2DEG. The bulk integer filling factor $\nu$ is deduced from the oscillations in the Hall bar. In the QPC trace, the more complex series of oscillations in the vicinity of the integer filling factor $\nu$ is highlighted in Fig. \ref{figure1}d. Similar resistance oscillations have been reported in QPCs and were attributed to the interaction between the counterpropagating QHECs through a small QHI (see Fig. \ref{figure1}c) located in the vicinity of the constriction \cite{Ford1994,Simmons1991,Hwang1991, Goldman1995,Goldman2008,Hackens2010,Martins2013}. The same mechanisms are at play in the present sample. We will indeed show below that successive types of QHIs, whose configuration depend on the magnetic field, are located in the constriction region. 

We first focus on the $R_{xx}$ evolution around $\nu=4$ (Fig. \ref{figure1}d) and discover four different QHI configurations as $B$ is raised (Figs. \ref{figure1}e-h). The whole picture, detailed hereafter, relies on a peculiar electrostatic potential landscape in the constriction region, with a central dot (larger electron density) surrounded by a region of lower electron density, fully determined by disorder-related potential fluctuations. This model is further justified in Appendix \ref{dot_antidot}, in particular with respect to another hypothesis involving a central antidot instead of a dot. 

At the onset of the $\nu=4$ QH state ($5.2 ~ {\rm T} \lesssim B \lesssim 5.6 ~ {\rm T}$), $R_{xx}$ shows series of narrow peaks superimposed on a $0 ~ \Omega$ plateau. This indicates the presence of a QHI formed as the counterpropagating QHECs (red channels at the top and bottom of Fig. \ref{figure1}e) are coupled at both sides of the QPC (red dotted lines in Fig. \ref{figure1}e).

A large $R_{xx}$ peak is then observed ($5.6 ~ {\rm T} \lesssim B \lesssim 6 ~ {\rm T}$). It is the signature of a large transmission probability between the counterpropagating QHECs, as sketched in Fig. \ref{figure1}f.

As $B$ is further increased ($6 ~ {\rm T} \lesssim B \lesssim 6.4 ~ {\rm T}$), $R_{xx}$ exhibits dips below a $\sim 2.2 ~ {\rm k\Omega}$ plateau (red dashed line in Fig. \ref{figure1}d). Such a plateau appears when the inner QHEC is completely backscattered so that the region outside of the QPC exhibits a different filling factor ($\nu$ in the bulk) from the QPC region (filling factor $\nu^*$), as illustrated in Fig. \ref{figure1}g. In such configuration, the resistance is given by \cite{Buettiker1988}
\begin{equation}\label{Eq1}
R_{xx}=\frac{h}{e^2}\left( \frac{1}{\nu^*}-\frac{1}{\nu} \right),
\end{equation}
where $e$ is the electron charge and $h$ is the Planck constant. The $2.2 ~ {\rm k\Omega}$ plateau is consistent with $\nu = 4$ and $\nu^* = 3$. It indicates that the spin degeneracy of the LLs has been lifted due to Zeeman effect \cite{Desrat2004}. The second LL (and its corresponding QHEC) is therefore split in two states: one spin down-polarized (in red in Figs. \ref{figure1}e-h) and one spin up-polarized (in orange). Based on this picture, the dips below the plateau are explained by the presence of a dot (QHEC loop) at the center of the constriction which acts as a QHI coupling the QHECs running at the left and right sides of the QPC in Fig. \ref{figure1}g. When the QHI formed by this dot (Fig. \ref{figure1}c) is active, these QHECs are not perfectly backscattered anymore and $R_{xx}$ drops below the $2.2 ~ {\rm k\Omega}$ plateau.

In the next $B$-range ($6.4 ~ {\rm T} \lesssim B \lesssim 6.7 ~ {\rm T}$), $R_{xx}$ peaks are observed above the $2.2 ~ {\rm k\Omega}$ plateau. This magnetoresistance sequence is close to the first regime (Fig. \ref{figure1}e) but the QHI is created by the spin up-polarized QHEC while the spin down-polarized QHEC is perfectly backscattered (Fig. \ref{figure1}h). 

When $B$ is further increased, the regimes highlighted in Fig. \ref{figure1}d are repeated with the spin up-polarized QHEC. $R_{xx}$ then features fluctuations around a plateau close to $4.3 ~ {\rm k\Omega}$ (right red dashed line in Fig. \ref{figure1}d), corresponding to the filling factors $\nu = 3$ and $\nu^* = 2$ in Eq. \ref{Eq1}. In the remainder of this paper, we will further characterize the QHI formed by the spin-up polarized QHEC in the configuration of Fig. \ref{figure1}g.

\begin{figure}[!ht]
\centering
\includegraphics[width=\linewidth]{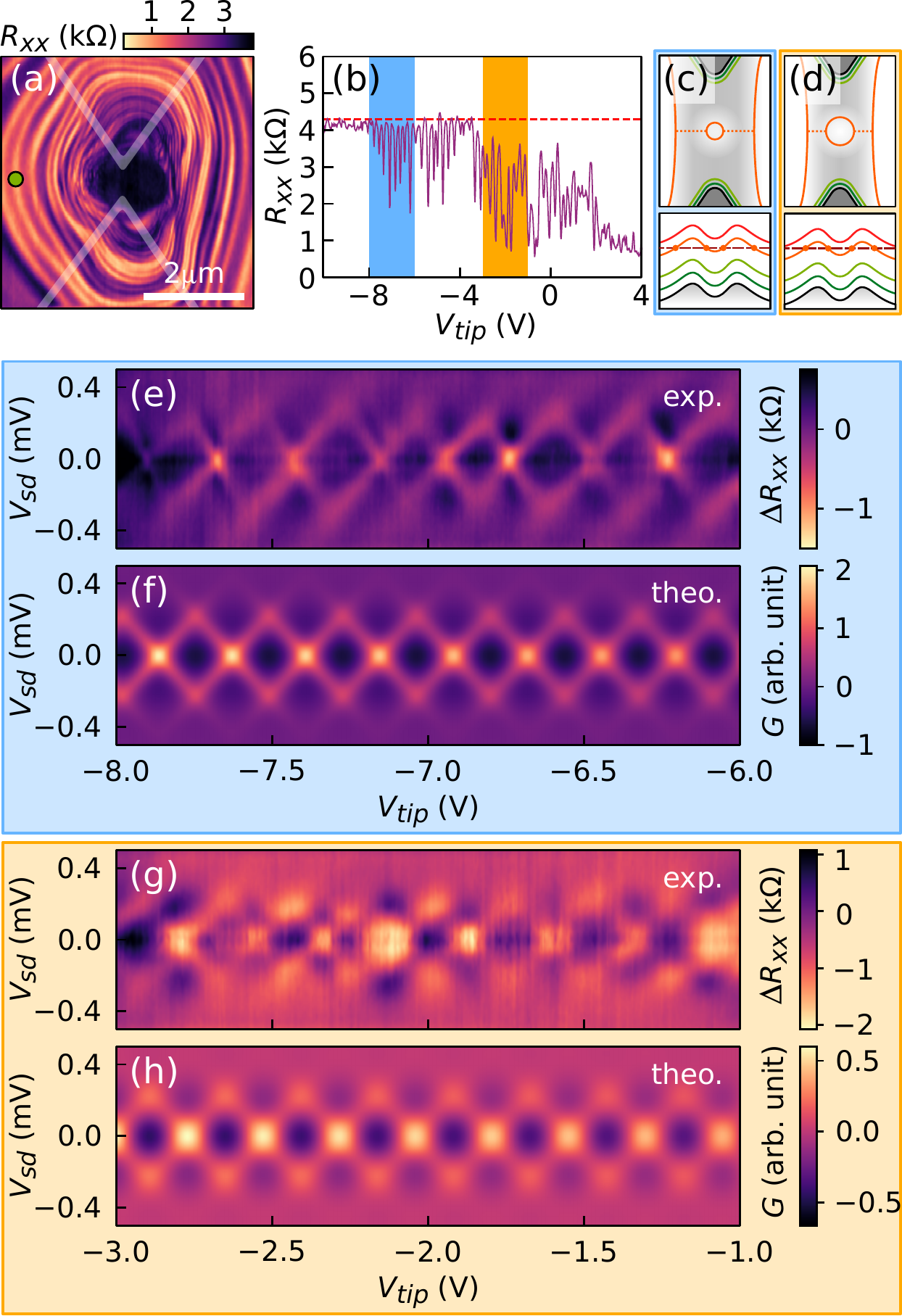}
\caption{(a) SGM map obtained by recording $R_{xx}$ as a function of the tip position at $B = 7.3$ T (blue dashed line in Fig. \ref{figure1}d) and with $V_{tip} = 0$ V. Note that due to the work function difference between tip and sample surface, $V_{tip} = 0$ V corresponds to a depleting tip. The borders of the sample are indicated by the white shaded area. (b) Evolution of $R_{xx}$ with $V_{tip}$ for the tip position indicated with a green dot in (a). The red dashed lines indicates the $4.3 ~ {\rm k\Omega}$ plateau. (c,d) Evolution of the QHECs spatial configuration with $V_{tip}$ in the vicinity of the constriction (top panels) and the corresponding Landau level positions compared to the Fermi energy $E_F$ (bottom panel) (c) and (d) correspond respectively to the blue and orange regions in (b). (e,f) Experimental map of $\Delta R_{xx}$ and simulated map of $G$, respectively, as a function of $V_{sd}$ and $V_{tip}$ (blue region in (b)). The maps feature diamond patterns. (g,h) Experimental map of $R_{xx}$ and simulated map of $G$, respectively, as a function of $V_{sd}$ and $V_{tip}$ (orange region in (b)). The maps feature checkerboard patterns. }
\label{figure2}
\end{figure}

\begin{figure}[!ht]
\centering
\includegraphics[width=\linewidth]{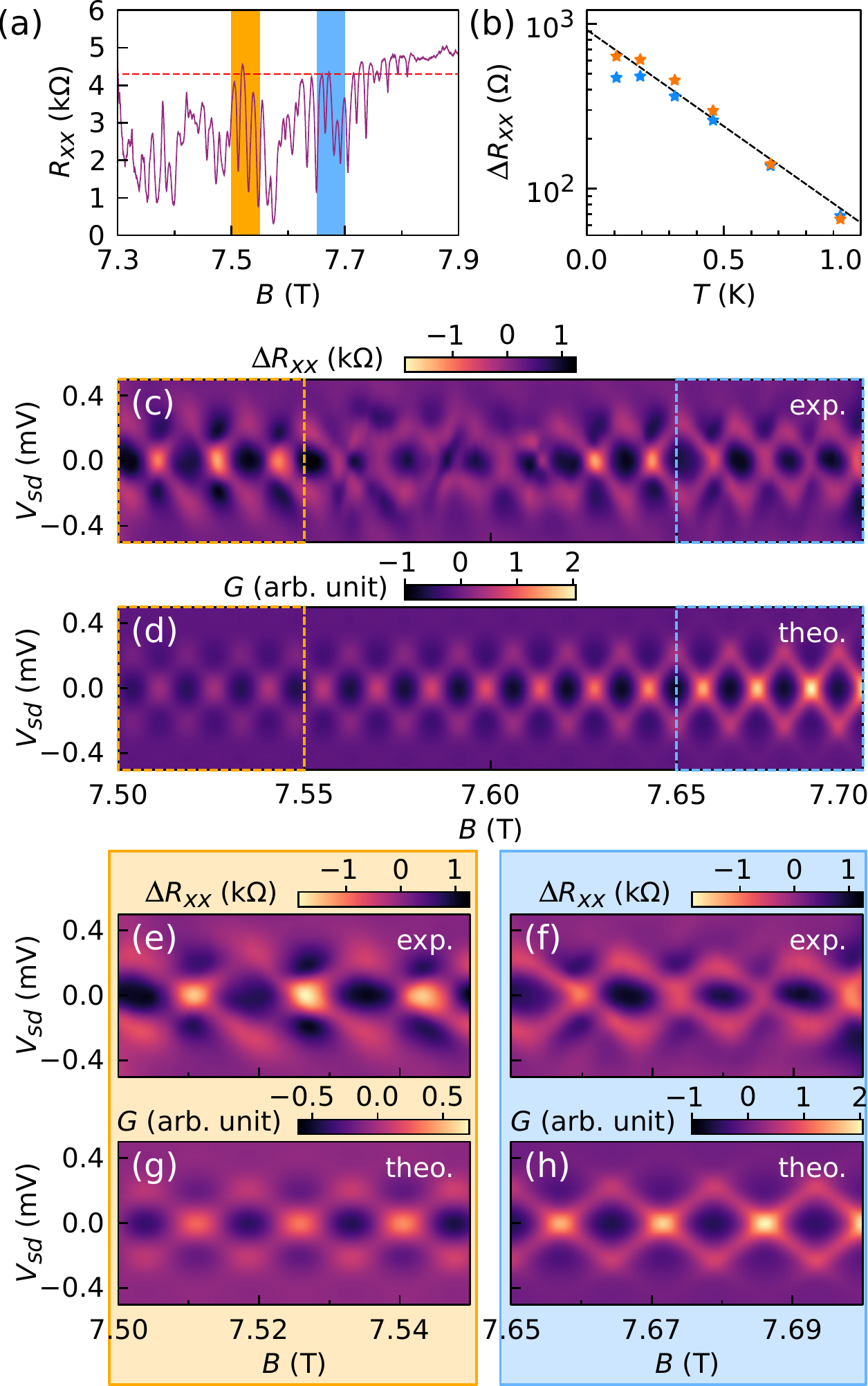}
\caption{(a) $R_{xx}$ vs $B$ (blue region in Fig. \ref{figure1}d) with $V_{tip} = 0$ V. The red dashed line indicates the $4.3 ~ {\rm k\Omega}$ plateau. (b) Oscillation amplitude $\Delta R_{xx}$ measured at different temperatures $T$ for the two ranges of $B$ (orange and blue regions in (a)). The temperature dependence, fitted with a continuous line, displays a similar exponential evolution for both $B$ ranges. (c,d) Experimental map of $\Delta R_{xx}$ and simulated map of $G$ as a function of $V_{sd}$ and $B$. (e,g) Zooms on the low $B$ range of (c,d) displaying a checkerboard pattern. (f,h) Zooms on the high $B$ range of (c,d) displaying a diamond pattern.}
\label{figure3}
\end{figure}

\subsection{DC bias spectroscopies}\label{spectro}
Figure \ref{figure2}a displays a $R_{xx}$ map as a function of the electrically-biased tip position, with the etched area defining the constriction indicated by the white shaded regions. This map features concentric fringes centered on the QPC, which clearly confirms that the QHI is located in the constriction. In the AB framework, these oscillations originate from a change of the magnetic flux $\Phi = BA$ enclosed in the QHI whose area $A$ varies when the electrically-biased tip moves in its vicinity. As a result, interference conditions in the QHI alternate between constructive and destructive, which yields concentric fringes in the SGM image. Note that another explanation could be given for the same phenomenon in the framework of the CD regime (with the tip influencing energy levels of a tunnel-coupled quantum dot), but we will see below that we can discard this hypothesis.

Figure \ref{figure2}b shows $R_{xx}$ as a function of $V_{tip}$ for the tip position indicated with a green dot in Fig. \ref{figure2}a. When $V_{tip}$ is tuned to large negative values (below $\sim -3$ V), $R_{xx}$ reaches the $4.3 ~ {\rm k\Omega}$ plateau, decorated with a set of narrow dips. This behavior is perfectly consistent with the picture presented in Fig. \ref{figure1}. Indeed, a negative $V_{tip}$ decreases the electron density (i.e., the electrostatic potential landscape shifts upwards) which implies a smaller QHI size. Consequently, the transmission probability to the backscattered QHECs is decreased (Fig. \ref{figure2}c) compared to a less negative $V_{tip}$ (Fig. \ref{figure2}d). With a smaller transmission probability, $R_{xx}$ remains close to the plateau that corresponds to a perfect backscattering of the QHECs.

Next, we compare the DC bias spectroscopies in the blue (weak transmission) and orange (stronger transmission) ranges in Fig. \ref{figure2}b, to obtain further information on the different regimes of transmission between the dot and the QHECs. In such measurements, one records $R_{xx}$ as a function of $V_{tip}$ and the source-drain voltage $V_{sd}$. Both parameters change the phase of the QHI. A high-pass filter has been applied to each map to get rid of the broad background evolution of $R_{xx}$ and only keep the oscillations $\Delta R_{xx}$ originating from the QHI. First, Fig. \ref{figure2}e shows the spectroscopy obtained at large negative values of $V_{tip}$ (weak transmission - blue in Fig. \ref{figure2}b). In this situation, $\Delta R_{xx}$ features a characteristic diamond pattern, usually associated with Coulomb charging effects \cite{Kataoka1999,McClure2009,Martins2013a}. Second, when examining data at less negative values of $V_{tip}$ (stronger transmission - orange in Fig. \ref{figure2}b), the obtained spectroscopy displays a checkerboard pattern (Fig. \ref{figure2}g). The latter pattern is the signature observed in QHIs in the AB regime \cite{Neder2006,Roulleau2007,McClure2009,Martins2013,Wei2017,Nakamura2019,Jo2021,Deprez2021,Ronen2021}. 

However, the apparent transition from AB to CD regime when varying $V_{tip}$ seems surprising for at least two reasons. First, the characteristic energy scales in $V_{sd}$ are expected to be different, because they are ascribed to different and uncorrelated mechanisms: a phase change explains the AB oscillations in $V_{sd}$ whereas the Coulomb charging energy should determine the height of the Coulomb diamonds. Interestingly, the energy scale extracted from the spectroscopies in Figs. \ref{figure2}e,g remains the same for the checkerboard and the diamonds, with a value around 0.2 mV. Second, the Coulomb charging energy is determined by the size of the QHI and we will show in the next section that this size does not change significantly within the $V_{tip}$ range of Fig. \ref{figure2}. Therefore, if the Coulomb charging energy does not dominate in Fig. \ref{figure2}g, nor should it in Fig. \ref{figure2}e. Figure \ref{figure3}c, obtained by varying $B$ instead of $V_{tip}$, confirms that the checkerboard pattern (zoom shown in Fig. \ref{figure3}e) evolves continuously, and with a constant energy scale in $V_{sd}$, to a diamond pattern (zoom shown in Fig. \ref{figure3}f) with a small increase of the magnetic flux enclosed in the QHI. 

Studying the evolution of $R_{xx}$ oscillation amplitude $\Delta R_{xx}$ with the temperature yields a third reason to discard the hypothesis of a different origin for checkerboard and diamond patterns. The oscillation amplitude associated with both patterns (orange and blue regions in Fig. \ref{figure3}a) turns out to follow the same exponential dependence $\Delta R_{xx} = \Delta R_{xx,0}\exp(-T/T_0)$ with the temperature $T$ (Fig. \ref{figure3}b), consistent with the AB effect \cite{Martins2013}. The CD regime, on the contrary, would imply a $T^{-1}$ dependence \cite{Hackens2010}. These observations suggest that the QHI remains in the pure AB regime, whatever the observed phenomenology (checkerboard- or diamond-like). In that framework, it remains to explain how a diamond pattern can be observed in the AB regime.

\section{Discussion}
To explain the transition between checkerboard and diamond patterns in Figs. \ref{figure2}c,f and \ref{figure3}c, we have considered that the QHI acts always as a simple FP interfometer in the AB regime whatever the $V_{tip}$ or $B$ range. From this hypothesis, we have developed a model inspired from ref. \cite{Deprez2021} and described in details in Appendix \ref{FP_appendix}. Our model relies on the schematic shown in Fig. \ref{figure1}c, where the FP QHI is formed by a closed-loop QHEC (inside a dot) connected to the left (input) and the right (output) QHECs by two beam splitters (dotted lines). 

Hereafter, we show that this model accurately reproduces the experimental spectroscopy results. Before developing the comparison between experimental and simulation results, we should stress that experiment yields maps of $\Delta R_{xx}$, the oscillating component of the resistance through the sample, whereas model results, obtained from Eq. \ref{G}, correspond to oscillating component of the conductance $G$ through the dot. The two quantities are directly linked since a large value of $G$ favors the transport through the QPC, lowering $R_{xx}$ below the plateau value, and therefore increasing $\Delta R_{xx}$. Note also that the spectroscopies shown in Figs. \ref{figure2}e,g and \ref{figure3}c feature a decrease of the oscillations visibility when $|V_{sd}|$ is raised. Nevertheless, the origin of the phase randomization mechanism leading to this behavior is debated in the literature and a consensual explanation is still missing \cite{Youn2008,Levkivskyi2008,Kovrizhin2009,Schneider2011}. Here, we have modeled the visibility decay with a Gaussian function (Eq. \ref{G}) fitted to the data.

The correspondence between the simulated $G$ and experimental $\Delta R_{xx}$ spectroscopies is illustrated in Figs. \ref{figure2}e,f and Figs. \ref{figure2}g,h. Both simulated maps were obtained with the same parameters in Eq. \ref{G}: $B = 7.3$ T as in the experiment, the QHEC velocity $v = 0.5\cdot 10^{5}$ m/s and the parameters associated to the visibility decay $\gamma = 1$ and $\Delta V_{sd}= 0.6$ mV. The following parameters were the only ones adjusted when changing $V_{tip}$ : $L = 404.6$ nm and $R_L = R_R = 0.05$ at $V_{tip} = -8$ V, and $L = 435.1$ nm and $R_L = R_R = 0.3$ at $V_{tip} = -1$ V (These values were used to find the free parameters in Eqs. \ref{LVtipB} and \ref{Rexp}). Three main outcomes can be highlighted from simulation results, and the good fit to the experimental data: (i) At large negative $V_{tip}$, a good correspondence has been obtained for a small dot size and a large reflection probability (weak transmission) and a diamond pattern is observed in this case, whereas at smaller negative $V_{tip}$ (larger dot, stronger transmission), the model yields a checkerboard pattern. (ii) The dot size obtained from a fit to the AB oscillation is fully consistent with the constriction width of 350 nm. Indeed, the dot diameter is around 250 nm. (iii) The QHEC velocity $v$ used to fit the $V_{sd}$ oscillations is fully consistent with values obtained in previous studies \cite{McClure2009,Martins2013}.

Figures \ref{figure3}d,g,h also show that the model can accurately reproduce the experimental data of Figs. \ref{figure3}c,e,f, respectively. The simulations have been performed with an evolution of the QHI parameters from $L = 427.3$ nm and $R_L = R_R = 0.3$ at $B = 7.5$ T to $L = 407.8$ nm and $R_L = R_R = 0.05$ at $B = 7.7$ T. The other constant parameters are the same as in Fig. \ref{figure2}. Again, the dot shape evolution and its coupling to QHECs follow the pictures proposed in Fig. \ref{figure1}g, with a size and a transmission probability decreasing with the magnetic field. This set of maps further draws a clear overview of the evolution from a checkerboard to a diamond pattern as the dot geometry and transmission are changed.

In Figs. \ref{figure2} and \ref{figure3}, we have accurately reproduced all experimental spectroscopy data with a model of QHI operating in the AB regime and whose radius is around 125 nm ($A < 0.05 ~ \mu \text{m}^2$). We have also observed QHIs functioning in the same regime with similar sizes in graphene \cite{Moreau2021c}. However, up to our knowledge, no QHI operating in the AB has ever been reported with a size under $1~\mu \text{m}^2$, even with architectures optimized to reduce Coulomb interactions \cite{Nakamura2019,Deprez2021,Ronen2021}. According to ref. \cite{Halperin2011}, finding a QHI in the AB or CD regime depends on the capacitive coupling between the QHEC loop forming the QHI and the localized states enclosed in the QHI, caused by potential fluctuations. When the QHI size decreases, the capacitive coupling is enhanced and Coulomb interactions tend to dominate the transport. The main difference between the configuration measured here and previous studies is the absence of localized states inside the QHEC loop forming the QHI: the quantum dot confinement results from a "simple" single dip in the potential landscape (Fig. \ref{figure1}g). This advantageous potential landscape is the key to reach a fully AB-dominated regime in nanoscale QHI.

\section{Conclusion}
In this article, we have used SGM to reveal the presence of a quantum dot located in the vicinity of a QPC. In the QH regime, it acts as a FP inteferometer that modulates electrons backscattering between the two sides of the QPC. DC bias-spectroscopies reveal a continuous transition from checkerboard to diamonds patterns. With a simple FP model, we have shown that both patterns can be explained in a fully AB (coherent) regime, contrary to the usual practice that associates diamonds with the CD regime. We associate the observed transition with a change of transmission probability between the central QHEC loop and the counterpropagating QHECs flowing at each side of the QPC. Finally, we point out that finding a QHI with a size of a few hundredth of nanometers in a pure AB regime is surprising since it should yield a large charging energy. We resolve this apparent contradiction by pointing that no localized state exists inside the QHI, lowering strongly the probability of emergence of Coulomb interactions, a behavior also observed in graphene \cite{Moreau2021c}. Our work therefore paves the way towards the design of nano-sized QHIs allowing to feature both a pure AB regime and a coherence length much larger than the interferometer size, two key ingredients in the progress towards anyonic braining and tests of quantum physics groundwork.

\begin{acknowledgments}
This work has been supported by FRFC grants no. 2.4.546.08.F and 2.4503.12, FNRS grant no. 1.5.044.07.F, by the FSR and ARC programs “Stresstronics” and "NATURIST", by BELSPO (Interuniversity Attraction Pole IAP-6/42) and by the PNANO 2007 program of the ANR (MICATEC project). B.H. (research associate) and N.M. (FRIA fellowship) acknowledge financial support from the F.R.S.-FNRS of Belgium.
\end{acknowledgments}

\appendix

\section{Dot or Antidot?}\label{dot_antidot}
In Figs. \ref{figure1}e-h, we have presented the QHECs configurations in the vicinity of the QPC that explains the transport curve of Figs. \ref{figure1}b,d. This relies on a global decrease of the electron density in the constriction region and a local increase of the density, due to potential inhomogeneities, forming a quantum dot in the center of the constriction. However similar transport data in the literature have been explained with the presence of an antidot near the QPC, associated with a local decrease of the electron density \cite{Ford1994,Goldman1995,Goldman2008,Hackens2010,Martins2013}. In this appendix, we justify why the latter model antidot), despite being simpler, is not consistent with the present data. 

\begin{figure}[!ht]
\centering
\includegraphics[width=\linewidth]{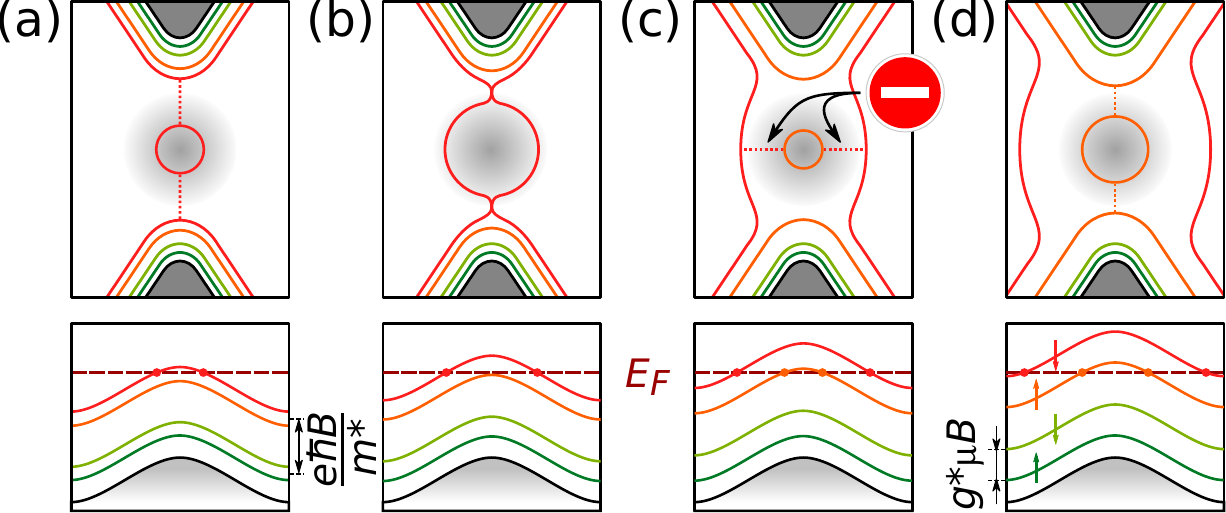}
\caption{(a-d) Evolution of the QHECs spatial configuration with $B$ in the vicinity an antidot located in the center of the constriction (top panels) and the corresponding Landau levels position compared to the Fermi energy $E_F$. A high potential (low electron density) is depicted with darker gray shades. }
\label{appendix1}
\end{figure}

Figure \ref{appendix1} presents the QHECs configurations in the situation where an antidot is located in the center of the constriction. At first sight, the four configurations are compatible with the different regimes in Fig. \ref{figure1}d. However, Fig. \ref{appendix1}c presents a first inconsistency. Indeed, we have shown that the spin degeneracy is lifted and the third (orange) and fourth (red) LLs are polarized with opposite spins. Therefore, the tunneling between the counterpropagating QHECs via the antidot in Fig. \ref{appendix1}c is prohibited. 

A second argument against the antidot model (and hence in favor of the model proposed in Figs. \ref{figure1}e-h) is found in the $R_{xx}$ evolution vs $V_{tip}$ in Fig. \ref{figure3}b. As discussed in the main text, large negative values of $V_{tip}$, leading to a decrease of the electron density, are associated with high reflection probabilities at the QHI beam splitters. It is inconsistent with the antidot model where a decrease of the electron density should yield a larger antidot, closer to the counterpropagating QHECs, leading to smaller reflection coefficient at the beam splitters (due to a higher transmission probability between the QHECs and the antidot). On the contrary, the model of Figs. \ref{figure1}e-h correctly captures the evolution of the transmission probability with $V_{tip}$.

\section{Fabry-Perot Model}\label{FP_appendix}
In this appendix, we develop the model used to produce the simulated maps of Figs. \ref{figure2} and \ref{figure3}. It relies on the schematic shown in Fig. \ref{figure1}c. In that framework, the wavefunction $\psi_D$ at the left side of the dot can be written
\begin{equation}\label{psiD}
    \psi_D = it_L\psi_{in} + r_Rr_L \exp(i2\phi_D)\psi_D,
\end{equation}
where $t_L$ is the transmission amplitude at the left side of the dot, $r_L$ and $r_R$ are the reflection amplitude at the left and right side of the dot, $\psi_{in}$ is the wavefunction of the left QHEC and $\phi_D$ is the phase accumulated along one arm of the QHI with a length $L= \sqrt{\pi A}$. Following the same logic, the wavefunction of the right QHEC is given by
\begin{equation}\label{psiout}
    \psi_{out} = it_R\exp(i\phi_D)\psi_D,
\end{equation}
where $t_R$ is the transmission amplitude at the right side of the dot. From Eqs. \ref{psiD} and \ref{psiout}, we define the transmission probability through the dot $T = |\psi_{out}|^2/|\psi_{in}|^2$ as a function of the reflection probabilities at each beam splitter $R_{L(R)} = |r_{L(R)}|^2 = 1-|t_{L(R)}|^2$ and we obtain
\begin{equation}\label{TD}
    T(\phi_D) = \dfrac{(1-R_L)(1-R_R)}{1+R_LR_R-2\sqrt{R_LR_R}\cos(\phi_D)}.
\end{equation}
This transmission depends on the phase $\phi_D$ that can be developed as \cite{C.Chamon1997}
\begin{equation}\label{phiD}
    \phi_D(\epsilon, \Phi) = \pi\frac{\Phi}{\Phi_0}+\frac{L\epsilon}{\hbar v} + \varphi,
\end{equation}
where the first term of the sum is the AB phase, with $\Phi_0 = h/2e$ the quantum of magnetic flux, the second term is the phase associated with the source-drain bias-induced energy shift $\epsilon$ where $v$ is the QHEC velocity in the QHI and the third term $\varphi$ is a constant phase. From Eq. \ref{TD}, one can extract the oscillating component of the transmission by subtracting the average term $<T(\epsilon, \Phi)>$. It yields
\begin{equation}
    T_{osc}(\epsilon, \Phi) = T(\epsilon, \Phi) - \frac{(1 - R_L)(1 - R_R)}{1 + R_LR_R}.
\end{equation}
The oscillation component of the current through the dot can then be expressed as
\begin{equation}\label{Iosc}
    I_{osc} = \frac{-e}{h}\int_{-eV_{sd}/2}^{eV_{sd}/2}T_{osc}(\epsilon, \Phi)d\epsilon,
\end{equation}
where $V_{sd}$ is the DC source-drain bias voltage. We assume an equal voltage drop at both beam splitters. The conductance $G$ through the QHI is obtained from Eq. \ref{Iosc} and the visibility decay is modeled with a Gaussian function in $V_{sd}$ as \cite{Roulleau2007,McClure2009,Martins2013}
\begin{equation}\label{G}
    G = \frac{dI_{osc}}{dV_{sd}} \exp\left(-2\pi\gamma\left(\frac{V_{sd}}{\Delta V_{sd}}\right)^2\right),
\end{equation}
where $\gamma$ and $\Delta V_{sd}$ are fitting parameters.

In the experimental spectroscopies of Figs. \ref{figure2} and \ref{figure3}, the AB phase evolves with $\Phi$ (Eq. \ref{phiD}) through $V_{tip}$ or $B$. Indeed, both parameters tune the dot area $A$ as illustrated in Figs. \ref{figure1}f-h for a change of $B$, modifying the spacing between the LLs, and in Figs. \ref{figure2}e,h for a change of $V_{tip}$, shifting the potential and the LLs compared to the Fermi energy $E_F$. The QHI size, given by the arm length $L$, then evolves as
\begin{equation}\label{LVtipB}
    dL = \frac{\partial L}{\partial V_{tip}}dV_{tip} + \frac{\partial L}{\partial B}dB.
\end{equation}
Since the terms $\partial L/\partial V_{tip}$ and $\partial L/\partial B$ are extremely difficult to assess, we assume that they are constant. The evolution of $L$ with $V_{tip}$ and $B$ is therefore linear.

Furthermore, any change of the dot size also affects its proximity, and hence its transmission probability (dotted lines in Figs. \ref{figure1}e-h and Figs. \ref{figure2}e,h), with the backscattered QHECs. Since the coupling occurs through tunneling, the associated transmission probability evolves exponentially with the distance between the dot and the QHECs. In turn, since this distance evolves linearly with the dot size, one has
\begin{equation}\label{Rexp}
    R_{L(R)} = \alpha \exp(-\beta L),
\end{equation}
where $\alpha$ and $\beta$ are positive coefficients to ensure that the reflection probability increases when the dot size decreases (see Fig. \ref{figure1}c). This expression is only valid when $R_{L(R)}\ll 1$.

\end{document}